\documentclass[12pt]{article}
\usepackage[utf8]{inputenc}
\usepackage{graphicx}
\usepackage{dcolumn}
\usepackage[style=numeric-comp, sorting=none]{biblatex}
\usepackage{array}
\usepackage{caption}
\usepackage[OT1]{fontenc}
\usepackage{amsfonts, amsmath, amsthm, amssymb}
\usepackage{listings}
\usepackage{xcolor}
\usepackage[margin=1.0in]{geometry}

\usepackage{authblk}

\newcolumntype{C}{>{\centering\arraybackslash}m{2em}}
\begin{document}

\title{Algebraic realisation of three fermion generations with $S_3$ family and unbroken gauge symmetry from $\mathbb{C}\ell(8)$}
\author[1]{Liam Gourlay}
\author[1]{Niels Gresnigt\thanks{niels.gresnigt@xjtlu.edu.cn}}
\affil[1]{Department of Physics, Xi’an Jiaotong-Liverpool University,\newline 111 Ren’ai Rd. Suzhou 215123, P.R. China}
\date{}

\maketitle

\begin{abstract}
Building on previous work, we extend an algebraic realisation of three fermion generations within the complex Clifford algebra $\mathbb{C}\ell(8)$ by incorporating a $U(1)_{em}$ gauge symmetry. The algebra $\mathbb{C}\ell(8)$ corresponds to the algebra of complex linear maps from the (complexification of the) Cayley-Dickson algebra of sedenions, $\mathbb{S}$, to itself. Previous work represented three generations of fermions with $SU(3)_C$ colour symmetry permuted by an $S_3$ symmetry of order-three, but failed to include a $U(1)$ generator that assigns the correct electric charge to all states. Furthermore, the three generations suffered from a degree of linear dependence between states. By generalising the embedding of the discrete group $S_3$, corresponding to automorphisms of $\mathbb{S}$, into $\mathbb{C}\ell(8)$, we include an $S_3$-invariant $U(1)$ that correctly assigns electric charge. First-generation states are represented in terms of two even $\mathbb{C}\ell(8)$ semi-spinors, obtained from two minimal left ideals, related to each other via the order-two $S_3$ symmetry. The remaining two generations are obtained by applying the $S_3$ symmetry of order-three to the first generation. In this model, the gauge symmetries, $SU(3)_C\times U(1)_{em}$, are $S_3$-invariant and preserve the semi-spinors. As a result of the generalised embedding of the $S_3$ automorphisms of $\mathbb{S}$ into $\mathbb{C}\ell(8)$, the three generations are now linearly independent.
\end{abstract}

\section{Introduction}

In the search for a minimal mathematical framework providing a derivation of the Standard Model (SM), the endeavour to link the four division algebras, and their left multiplication algebras to particle physics has gained substantial traction in recent years, with numerous lecture series on the subject \cite{FureyLectures1,FureyLectures2,OSMU2023, BoyleLectures} and a proliferation of research papers
\cite{Wilson2022,Furey2023One,Furey2023Two, Furey2023Three,Todorov2022,todorov2020superselection,gresnigt2020topological, gresnigt2021topological,gresnigt2018braids,Furey2022a,Furey2022,Gording2019,Krasnov2021,Krasnov2022,Lu2024,Masi2021,Stoica2017,Perelman2019,Vaibhav2021,Furey2016,Furey2018,Furey2018a, singh2021quantum, finster2024causal}. In many such approaches, the gauge groups, leptons, and quarks are contained within the multiplication algebras. Although the composition of division algebras need not be associative (nor alternative), their multiplication algebras, the algebras generated from the actions of division algebras on itself via its endomorphisms, is associative.

There are four normed division algebras; $\mathbb{R}$, $\mathbb{C}$, $\mathbb{H}$ (quaternions) and $\mathbb{O}$ (octonions), of dimensions one, two, four and eight, respectively. Shortly after the discovery of quarks, G\"uynadin and G\"ursey \cite{Gunaydin1973} constructed a model of quark colour symmetry based on the algebra of the split octonions\footnote{It is important to note that the construction is possible only in the split octonions, and not in the octonions themselves.}. Subsequently, a series of papers by Barducci et al, and Casalbuoni and Gatto \cite{Barducci1977, Casalbuoni1979, Casalbuoni1980} explored a unified description of leptons and quarks with internal degrees of freedom in terms of fermionic oscillators (corresponding to a Witt basis of a Clifford algebra, see section 2.3). Three fermionic oscillators\footnote{The abstract algebra of three fermionic oscillators has two different realisations, one in terms of split octonions \cite{Gunaydin1973}, the other in terms of the (associative) complex Clifford algebra $\mathbb{C}\ell(6)$.} are associated with the colour degrees of freedom. The inclusion of electric charge requires a fourth fermionic oscillator (which in \cite{Barducci1977, Casalbuoni1979, Casalbuoni1980} is unrelated to division algebras).

The early association of the split octonions with quarks in \cite{Gunaydin1973} was expanded upon by Dixon \cite{Dixon1990, Dixon2004, Dixon1994, Dixon2010} who revealed that the mathematical characteristics of the SM, encompassing its gauge symmetries and corresponding multiplets to which a single generation of fermions is subject, are inherent in $\mathbb{T}^2$, where $\mathbb{T}=\mathbb{R}\otimes\mathbb{C}\otimes\mathbb{H}\otimes\mathbb{O}$, commonly referred to as the Dixon algebra. Here, $\mathbb{T}^2$ corresponds to a complexified (hyper) spinor in 1+9D spacetime.

In an approach closely related to these earlier works \cite{Gunaydin1973, Barducci1977, Casalbuoni1979, Casalbuoni1980, Dixon1990, Dixon2004, Dixon1994, Dixon2010}, Furey encompasses both bosons and fermions within Clifford algebras, arising as the multiplication algebras of compositions of division algebras. Two minimal left ideals of $\mathbb{C}\ell(6)$, the left (or right) multiplication algebra of $\mathbb{C}\otimes\mathbb{O}$, transform as a single generation under $SU(3)_C\times U(1)_{em}$, whereas two $\mathbb{C}\ell(4)$ minimal ideals, which can be generated from the left and right multiplication algebra of $\mathbb{C}\otimes\mathbb{H}$, transform as a single generation of chiral fermions under weak $SU(2)$ \cite{Furey2018b}. These findings integrate into a $\mathbb{C}\ell(10)$ model \cite{Gresnigt2020,Furey2023Three,todorov2020superselection}, whose $Spin(10)$ group generated from the bivectors of $\mathbb{C}\ell(10)$ can be systematically broken by requiring invariance under a series of division algebraic reflections, leading to a cascade of grand unified theories (GUTs) \cite{Furey2023One, Furey2023Two, Furey2023Three}.

The representation of spinors as minimal left ideals dates back to the 1930s \cite{Juvet1930,Sauter1930} and 1940s \cite{riesz2013clifford}. In Furey's model, fermions are identified with the basis states of minimal left ideals in the multiplication algebra. This association is established through a standard procedure utilising a Witt decomposition, as reviewed by Ab\l amowicz \cite{Ablamowicz1995}. The gauge symmetries within this framework are characterised as those unitary symmetries that preserve these minimal left ideals under commutation.

Most division-algebraic based constructions to date are limited to describing a single generation of fermions. Despite various attempts \cite{Furey2018a,Furey2014,dixon2014,Gillard2019}, an algebraic foundation for the existence of three generations within the division algebraic framework remains elusive\footnote{The most popular GUTs, such as those based on $SU(5)$, $SO(10)$, and the Pati-Salam model are also inherently single-generation models, lacking a theoretical basis for three generations.}. Furey endeavours to depict three generations directly from the algebra $\mathbb{C}\ell(6)$ \cite{Furey2018a,Furey2014}. After defining two copies of $SU(3)$, the remaining 48 degrees of freedom are found to transform as three generations of colour states. The states are then, however, no longer described in terms of minimal left ideals, as is the case for a single generation \cite{Furey2016}. Additionally, the $U(1)_{em}$ generator (corresponding to the number operator), which assigns the correct electric charge in the context of a one-generation model, fails to work in this three-generation model, although a modified construction allows for $U(1)_{em}$ to be included \cite{Furey2018a}. A similar construction based on $\mathbb{C}\ell(6)$, which includes an $SU(2)$ gauge symmetry, is given by Gording and Schmidt-May \cite{Gording2019}. Dixon, on the other hand, characterises three generations using the algebra $\mathbb{T}^6=\mathbb{C}\otimes\mathbb{H}^2\otimes\mathbb{O}^3$ \cite{Dixon2004}. The choice of $\mathbb{T}^6$, over other $\mathbb{T}^{2n}$ options, appears arbitrary but can be motivated from the Leech lattice. In the unified theories of \cite{Casalbuoni1980}, an additional $m$ fermionic oscillators are included in order to represent $2^m$ generations (a necessarily even number). However, these additional fermionic oscillators cannot be associated with the division algebras in an obvious way, nor is there any algebraic guidance on what $m$ should be. Other authors have sought to encode three generations within the exceptional Jordan algebra $J_3(\mathbb{O})$, comprising three-by-three matrices over $\mathbb{O}$ \cite{dubois2016exceptional1,dubois2019exceptional2,todorov2018octonions,todorov2018deducing,boyle2020standard2,boyle2020standard}. Each octonion is associated with one generation through the three canonical $J_2(\mathbb{O})$ subalgebras of $J_3(\mathbb{O})$.


The three-generation model proposed here is fundamentally different from the aforementioned ones. Existing models inevitably consider compositions (tensor products) of division algebras, which themselves are no longer division algebras. One might, therefore, ask if going beyond the division algebras and considering larger Cayley-Dickson algebras might provide additional algebraic structure suitable for describing three generations. The first Cayley-Dickson algebra beyond the octonions is the algebra of sedenions $\mathbb{S}$. The automorphism group of this algebra is $Aut(\mathbb{S})=Aut(\mathbb{O})\times{S_3}$. It is the appearance of the discrete group $S_3$ that motivates us to consider this algebra as a basis for constructing a three-generation model, and at the core of our construction is the idea that $S_3$ serves as the algebraic source for the existence of exactly three generations. The associative left multiplication algebra that is generated from $\mathbb{C}\otimes\mathbb{S}$ is the algebra $\mathbb{C}\ell(8)$, which will serve a central purpose in our construction. 

An initial attempt at a three-generation model in \cite{Gillard2019} used three $\mathbb{O}$ subalgebras of $\mathbb{S}$ to represent three generations, generalising the earlier three-generation lepton model using three $\mathbb{H}$ subalgebras of $\mathbb{O}$ \cite{Manogue1998}. However, this model had two drawbacks: each generation required its own copy of $SU(3)_C$ (suggesting three generations of gluons), and the $S_3$ automorphisms of $\mathbb{S}$ lacked a clear physical interpretation.

The more recent model \cite{Gresnigt2023}\footnote{See also the related works \cite{gresnigt2023sedenion,gresnigt2023toward}.} addressed these issues by utilising the order-three $S_3$ automorphism of $\mathbb{S}$ to generate three generations. Instead of associating three $\mathbb{O}$ subalgebras with three generations, the entire sedenion algebra was used to construct a single generation of colour states, with the $S_3$ automorphism of order-three generating two additional copies. This provides a clear physical interpretation for the order-three $S_3$ automorphism and keeps $SU(3)_C$ invariant under $S_3$, thereby avoiding three generations of gluons.

However, the updated model did not include a $U(1)$ symmetry that assigns the correct electric charge to states. Also, while the $S_3$ symmetry of order-three was given a clear physical interpretation, the order-two symmetry was not. Furthermore, in both models, the three generations of fermions were not linearly independent.

Here we overcome these limitations by including an $S_3$-invariant $U(1)$ symmetry that assigns the correct electric charge to all three generations of states. This is achieved by generalising the embedding of the $S_3$ automorphisms of $\mathbb{S}$ into $\mathbb{C}\ell(8)$. We initially represent one generation of electrocolour states in terms of a single $\mathbb{C}\ell(8)$ minimal left ideal\footnote{The minimal left ideal is invariant under an $SU(4)$ gauge symmetry. Restricting ourselves to the symmetry that leaves invariant a $\mathbb{C}\ell(2)$ subalgebra (equivalently, a quaternionic structure), breaks the symmetry to the maximal subgroup $SU(3)\times U(1)$.}. To generalise our construction to three generations, the generalised $S_3$ symmetry of order-three is applied to the minimal left ideal. Being invariant under $S_3$, the $SU(3)$ symmetry transforms all states correctly. The $U(1)$ generator, on the other hand, fails to assign the correct electric charge because it is not invariant under $S_3$. 

To include a $U(1)$ symmetry, we instead split each spinor into its even and odd-grade semi-spinors via a projector. Subsequently, we apply the order-two $S_3$ symmetry to the even semi-spinor, which results in a second even semi-spinor belonging to a different minimal left ideal. Applying the generalised order-three $S_3$ symmetry generates two additional pairs of semi-spinors. We then identify the gauge symmetries of our model with those unitary symmetries that both preserve the semi-spinors, and are invariant under this $S_3$ action. The required $S_3$-invariant $U(1)$ symmetry then arises as the sum of three individual $U(1)$ symmetries, and the three pairs of semi-spinors transform as three generations of fermions under the SM unbroken gauge symmetry $SU(3)_C\times U(1)_{em}$. As a byproduct, all three generations are now linearly independent.

\section{$\mathbb{C}\ell(8)$ as the left multiplication algebra of $\mathbb{C}\otimes\mathbb{S}$}

\subsection{Normed division algebras}

A division algebra is an algebra over a field where division is always well-defined, except by zero. A normed division algebra is a division algebra equipped with a norm, defined in terms of a conjugate. Hurwitz \cite{HurwitzUeberDC} showed that there are only four normed divisional algebras over the field of real numbers; $\mathbb{R}$, $\mathbb{C}$, $\mathbb{H}$ (quaternions) and $\mathbb{O}$ (octonions), of dimensions one, two, four and eight, respectively. Starting from the real numbers $\mathbb{R}$, the Cayley-Dickson (CD) construction produces a sequence of algebras, $\mathbb{A}_n$ (where $\mathbb{A}_0=\mathbb{R}$), of dimension $2^n$, the first three being the remaining three division algebras. 

The octonions $\mathbb{O}$ are the largest normed division algebra with seven mutually anti-commuting imaginary units $u_i$ ($i = {1,...,7}$), together with the identity $u_0$. A general octonion $x$ can be written as;
\begin{equation}
    x = x_0u_0 + x_1u_1+...+x_7u_7, \hspace{1em} x_0,...,x_7 \in \mathbb{R}.
\end{equation}

The multiplication of two octonion elements is calculated as $u_iu_j=-\delta_{ij}+f_{ijk}u_k$ for $i,j,k \in \{1,2,...,7\}$, where $f_{ijk}$ is an anti-symmetric tensor, $f_{ijk}=1$ when $ijk \in \{123, 145, 176, \allowbreak 246, 257, 347, 365\}$\footnote{This is not a unique choice and different authors use different multiplication tables.}, and zero otherwise. The standard involution of an octonion element $x$ is given by $\overline{x}=x_0u_0-x_1u_1-...-x_7u_7$. The norm $|x|$ is defined by $|x|^2=x\overline{x}=\overline{x}x$, and the inverse of $x$ is $x^{-1}={\overline{x}}/{|x|^2}$.

Elements $q_i$, $q_j$ and $q_k\in\mathbb{O}$ satisfying $q_iq_j=q_k$ correspond to a quaternion subalgebra of $\mathbb{O}$. There are seven such subalgebras, with elements as in $f_{ijk}$. Multiplication of octonion elements not belonging to the same quaternion subalgebra is non-associative. The automorphism group of $\mathbb{O}$ is the exceptional Lie group $G_2$. This exceptional group contains $SU(3)$ as one of its maximal subgroups, corresponding to the stabiliser subgroup of one of the octonion imaginary units. $SU(3)$ is thus the group that preserves a complex structure. It is this observation that lead G\"unaydin and G\"ursey \cite{Gunaydin1973} to relate the quark colour symmetry to the split octonions.

For a more detailed overview of division algebras, see \cite{baez2002octonions}.

\subsection{Sedenions, and the left multiplication algebra of $\mathbb{C} \otimes \mathbb{S}$}

Applying the CD construction to $\mathbb{O}$ generates the 16-dimensional algebra of sedenions $\mathbb{S}$. This algebra is non-commutative, non-associative, non-alternative, and contains zero divisors. An orthonormal basis consists of 15 imaginary units $s_i$ ($i = {1,...,15}$), as well as the identity $s_0$. The imaginary units $s_1,...,s_7$ correspond to the octonion imaginary units $u_1,...,u_7$. A general sedenion $w$ may then be written as;
\begin{equation}
    w = w_0s_0+w_1s_1+...+w_{15}s_{15}, \hspace{1em} w_0,...,w_{15} \in \mathbb{R}.
\end{equation}

The product of two sedenions can be determined using the multiplication table in Appendix A\footnote{More details on the sedenions and their properties can be found in \cite{Cawagas2004}.}. An example of two non-zero sedenions that multiply to zero is; $(s_1+s_{10})\cdot(s_5+s_{14})=0$.

The involution of a sedenion element $w$ is given by $\overline{w}=w_0s_0-w_1s_1-...-w_{15}s_{15}$. The norm $|w|$ is defined by $|w|^2=w\overline{w}=\overline{w}w$ and the inverse of $w$ (if it exists) is $w^{-1}={\overline{w}}/{|w|^2}$. 

It is known that the automorphism group of $\mathbb{A}_4 = \mathbb{S}$ is $Aut(\mathbb{S})=Aut(\mathbb{O})\times{S_3}$ \cite{Brown1967}. The automorphisms can be explicitly stated as follows;
\begin{align} \label{eqn: phi}
    \phi: A+Bs_8 &\rightarrow \phi(A)+\phi(B)s_8, \\ \label{eqn: epsilon}
    \epsilon: A+Bs_8 &\rightarrow A-Bs_8, \\ \nonumber
    \psi: A+Bs_8 &\rightarrow \frac{1}{4}[A+3A^*+\sqrt{3}(B-B^*)] \\ \label{eqn:Psi}
    &+\frac{1}{4}[B+3B^*-\sqrt{3}(A-A^*)]s_8,
\end{align}
where $A,B \in \mathbb{O}$. $A^*$ is an octonion involution of $A$ such that $(A^*)^*=A$ and $(AB)^*=B^*A^*$, and $\phi$ is an element of $G_2$, the automorphism group of $\mathbb{O}$. These result in the following identities; $\epsilon^2=Id$, $\psi^3=Id$, $\psi\phi=\phi\psi$, $\epsilon\phi=\phi\epsilon$ and $\epsilon\psi=\psi^2\epsilon$. It follows that $\epsilon$ and $\psi$ generate $S_3$.

Although $\mathbb{S}$ is both non-associative and non-alternative, it is still possible to define compositions of left or right actions of $\mathbb{S}$ on itself as linear operators, thereby generating an associative algebra \cite{lohmus1994}. This process mirrors the construction of $\mathbb{C}\ell(6)$ as the left (or right) multiplication algebra of the complex octonions, $\mathbb{C}\otimes\mathbb{O}$ \cite{Furey2016, Dixon1994, lohmus1994}. In this paper, we restrict our attention to the left associative multiplication algebra.

Let $L_a$ denote the linear operator corresponding to left multiplication by an element $a \in \mathbb{C} \otimes \mathbb{S}$ onto an element $w \in \mathbb{C} \otimes \mathbb{S}$, defined by;
\begin{equation}
    L_a[w]:= aw, \hspace{1em} \forall a,w \in \mathbb{C} \otimes \mathbb{S}.
\end{equation}

Since $L_a$ corresponds to a linear operator, it can be represented as a $16 \times 16$ complex matrix (acting on the vector space $\mathbb{C} \otimes \mathbb{S}$ written as a $16 \times 1$ column vector). Due to the non-associativity of $\mathbb{S}$, the left multiplication algebra of $\mathbb{C} \otimes \mathbb{S}$ contains new maps which are not captured by $\mathbb{C} \otimes \mathbb{S}$, because in general;
\begin{alignat}{4}
    L_aL_b[w] &= a(bw) &&\neq L_{ab}[w]&&&=(ab)w,\quad a,b,w\in\mathbb{C} \otimes \mathbb{S}.
\end{alignat}

There are a total of 256 distinct left-acting complex-linear maps from $\mathbb{C} \otimes \mathbb{S}$ to itself, and these provide a faithful representation of $\mathbb{C}\ell(8)$. It can be shown that $L_{s_i}L_{s_j}[w]=-L_{s_j}L_{s_i}[w]$ and $L_{s_i}L_{s_i}[w]=-w$, $i,j = 1,...,8$, $i \neq j$. As a result, the left multiplication actions $L_{s_i}$, $i=0,1,...,8$, are a generating basis for $\mathbb{C}\ell(8)$. From now on, the left actions will be denoted simply by $e_i:=L_{s_i}$ and assumed to be acting on arbitrary $w \in \mathbb{C}\otimes{\mathbb{S}}$. That is, instead of writing $L_{s_i}L_{s_j}w$, we simply write $e_ie_j$. The maps $e_k$, $k=9,...,15$, can then be expressed in terms of these $e_i$, $i=1,...,8$. For example\footnote{Expressions for all $e_k$, $k=9,...,15$ as $\mathbb{C}\ell(8)$ elements can be found in \cite{Gresnigt2023}.};

\begin{equation} \label{eqn:e9}
    {e_9} = -\frac{1}{2}{e_1e_2e_3e_4e_5e_8}+\frac{1}{2}{e_1e_2e_3e_6e_7e_8}+\frac{1}{2}{e_1e_4e_5e_6e_7e_8}-\frac{1}{2}{e_1e_8}.
\end{equation}

\subsection{Constructing a minimal left ideal of $\mathbb{C}\ell(8)$}

A general construction for creating spinor spaces as minimal left ideals of Clifford algebras is reviewed in \cite{Ablamowicz1995}. The process for $\mathbb{C}\ell(2n)$ is particularly simple. A Witt basis can be formed from the algebra's canonical orthonormal basis vectors, which are then used to create primitive idempotents on which the Clifford algebra is left multiplied. For $\mathbb{C}\ell(8)$, we define the Witt basis;

\begin{equation}
\begin{alignedat}{3}
    a_1 &:= \frac{1}{2}(-e_1+ie_5), \hspace{2em} a_1^\dagger &&:= \frac{1}{2}(e_1+ie_5), \\ 
    a_2 &:= \frac{1}{2}(-e_2+ie_6), \hspace{2em} a_2^\dagger &&:= \frac{1}{2}(e_2+ie_6), \\ 
    a_3 &:= \frac{1}{2}(-e_3+ie_7), \hspace{2em} a_3^\dagger &&:= \frac{1}{2}(e_3+ie_7), \\ 
    a_4 &:= \frac{1}{2}(-e_4+ie_8), \hspace{2em} a_4^\dagger &&:= \frac{1}{2}(e_4+ie_8).
\end{alignedat}
\end{equation}

These are fermionic oscillators that generate two totally isotropic subspaces, and satisfy the fermionic anticommutation relations;
\begin{equation} \label{eqn: ACRs}
\{a_i,a_j\} = \{a_i^\dagger, a_j^\dagger\} = 0, \hspace{1em} \{a_i,a_j^\dagger\} = \delta_{ij}.
\end{equation}

The nilpotents $\Omega_1=a_1a_2a_3a_4$ and $\Omega_1^\dagger=a_4^{\dagger}a_3^{\dagger}a_2^{\dagger}a_1^{\dagger}$ can be combined into a primitive idempotent; $v_1:=\Omega_1\Omega_1^\dagger$, physically representing the vacuum state. A minimal left ideal is then;
\begin{equation}
    \mathbb{C}\ell(8)v_1.
\end{equation}

Explicitly, we can write the 16 complex-dimensional ideal as follows;
\begin{equation} 
T_1 = (r_0+r_ja_j^\dagger+r_{jk}a_j^\dagger{a_k^\dagger}+r_{jkl}a_j^\dagger{a_k^\dagger}a_l^\dagger+r_{1234}a_1^\dagger{a_2^\dagger}a_3^\dagger{a_4^\dagger})v_1,
\end{equation}
where $j,k,l \in \{1,2,3,4\}$, $j \neq k \neq l$ and $r_0,r_j,r_{jk},r_{jkl},r_{1234}$ are complex coefficients. $T_1$ contains subspaces of different grades, determined by the number of $a_i^\dagger$ operators multiplied onto the primitive idempotent. Notably, $T_1$ can be expressed as $(\mathbb{C}\ell(6) \hspace{.1em} \oplus \hspace{.1em} \mathbb{C}\ell(6)a_4^\dagger)v_1$, where $a_k$ and  $a_k^\dagger$, $k=1,2,3$, constitutes a Witt basis for $\mathbb{C}\ell(6)$. This splitting can be obtained by applying the projectors $\eta_{\pm}=\frac{1}{2}(1\pm{ie_4e_8})$ to $T_1$, with $\eta_-T_1=\mathbb{C}\ell(6)v_1$ and $\eta_+T_1=\mathbb{C}\ell(6)a_4^\dagger{v_1}$.

\section{One fermion generation with $SU(3)\times U(1)$ gauge symmetry}

We now consider the transformations of the minimal left ideal basis states. Although any transformation of the form
\begin{eqnarray}\label{spinoraction}
  e_i\mapsto e^{i\phi_kg_k}e_ie^{-i\phi_kg_k},\quad \phi_k \in \mathbb{R},\quad g_k \in \mathbb{C}\ell(8), 
\end{eqnarray}
will preserve the anticommutation relations in eqn. (\ref{eqn: ACRs}), not all such transformations preserve the Witt basis or, equivalently, the isotropic subspaces generated by them. In particular, not all elements of $Spin(8)$, whose Lie algebra is generated by the bi-vectors of $\mathbb{C}\ell(8)$, leave these isotropic subspaces invariant. Imposing the additional restrictions that
\begin{eqnarray}
   [g_k, \sum_{i} \kappa_ia_i]=\sum_{j} \mu_ja_j,\quad \textrm{and}\quad [g_k, \sum_{i} \kappa'_ia_i^\dagger]=\sum_{j} \mu'_ja_j^\dagger,
\end{eqnarray}
and that transformations on $a_i^{\dagger}$ ($a_i$) commute with hermitian conjugation $\dagger$;
\begin{eqnarray}
   e^{i\phi_kg_k} a_i^\dagger e^{-i\phi_kg_k}=(e^{-i\phi_kg_k})^\dagger a_i^\dagger (e^{i\phi_kg_k})^\dagger,
\end{eqnarray}
reduces $Spin(8)$ to its maximal subgroup $U(4)=SU(4)\times U(1)$\footnote{See \cite{Furey2016} for a more detailed exploration of Witt basis preserving symmetries.}. These unitary symmetries preserve the Witt basis (and hence the minimal left ideals) via the action in eqn. (\ref{spinoraction}).

\begin{equation}
\label{eq:SU(4)}
\begin{alignedat}{3} 
    \Lambda_1 &=  -a_2^\dagger{a_1}-a_1^\dagger{a_2}, \hspace{1em} 
    \Lambda_2 &&= ia_2^\dagger{a_1}-ia_1^\dagger{a_2}, \hspace{1em} \Lambda_3 = a_2^\dagger{a_2}-a_1^\dagger{a_1}, \\
    \Lambda_4 &= -a_1^\dagger{a_3}-a_3^\dagger{a_1}, \hspace{1em} 
    \Lambda_5 &&= -ia_1^\dagger{a_3}+ia_3^\dagger{a_1}, \hspace{1em}
    \Lambda_6 = -a_3^\dagger{a_2}-a_2^\dagger{a_3}, \\ 
    \Lambda_7 &= ia_3^\dagger{a_2}-ia_2^\dagger{a_3}, \hspace{1em}
     \Lambda_8 &&= -\frac{1}{\sqrt{3}}(a_1^\dagger{a_1}+a_2^\dagger{a_2}-2a_3^\dagger{a_3}),  \\
         \Lambda_9 &= -a_4^\dagger{a_1}-a_1^\dagger{a_4}, \hspace{1em}    
    \Lambda_{10} &&= ia_4^\dagger{a_1}-ia_1^\dagger{a_4}, \hspace{1em} 
    \Lambda_{11} = -a_4^\dagger{a_2}-a_2^\dagger{a_4}, \\
    \Lambda_{12} &= ia_4^\dagger{a_2}-ia_2^\dagger{a_4}, \hspace{1em} 
    \Lambda_{13} &&= -a_4^\dagger{a_3}-a_3^\dagger{a_4}, \hspace{1em}
    \Lambda_{14} = ia_4^\dagger{a_3}-ia_3^\dagger{a_4}, \\ 
    \Lambda_{15} &= -\frac{1}{\sqrt{6}}(a_1^\dagger{a_1}+a_2^\dagger{a_2}+&&a_3^\dagger{a_3}-3a_4^\dagger{a_4}).
\end{alignedat}
\end{equation}

The $SU(4)$ generators, $\{\Lambda_1,...,\Lambda_{15}\}$, transform the 16 basis states of $T_1$ as $1\oplus4\oplus6\oplus\overline{4}\oplus1$ under commutation. $SU(4)$ can be broken to $SU(3)\times{U(1)}$\footnote{$SU(3)\times{U(1)}$ corresponds to the subgroup of $SU(4)$ that commutes with the projectors $\eta_{\pm}$. Another way to look at this is that $SU(3)\times{U(1)}$ corresponds to the subgroup of $SU(4)$ that commutes with the quaternionic structure generated by $e_4$ and $e_8$.}, where the $SU(3)$ generators are $\{\Lambda_1,...,\Lambda_8\}$. The $SU(3)$ and $U(1)$ generators are the generators that preserve both $\mathbb{C}\ell(6)v_1$ and $\mathbb{C}\ell(6)a_4^\dagger{v_1}$ individually. Under the action of $SU(3)$, the basis elements of both $\mathbb{C}\ell(6)v_1$ and $\mathbb{C}\ell(6)a_4^\dagger{v_1}$ transform as $1\oplus{3}\oplus\overline{3}\oplus{1}$, matching that of a single generation of fermions under $SU(3)_C$. Two examples of $T_1$ basis elements transforming under $\Lambda_1$ can be seen below;
\begin{equation}
\begin{aligned}\relax
    [\Lambda_1, a_1^\dagger v_1] &= (-a_2^\dagger a_1 - a_1^\dagger a_2 )(a_1^\dagger v_1)-(a_1^\dagger v_1 )(-a_2^\dagger a_1 - a_1^\dagger a_2 ) \\
    &= -a_2^\dagger a_1 a_1^\dagger v_1 \\
    &= -a_2^\dagger(1-a_1^\dagger a_1 )v_1 \\
    &= -a_2^\dagger v_1, \\
   [\Lambda_1, a_1^\dagger a_2^\dagger a_3^\dagger a_4^\dagger v_1] &= (-a_2^\dagger a_1 - a_1^\dagger a_2 )(a_1^\dagger a_2^\dagger a_3^\dagger a_4^\dagger v_1)-(a_1^\dagger a_2^\dagger a_3^\dagger a_4^\dagger v_1)(-a_2^\dagger a_1 - a_1^\dagger a_2 ) \\
    &= (a_1 a_2^\dagger + a_2 a_1^\dagger)(a_1^\dagger a_2^\dagger a_3^\dagger a_4^\dagger v_1) \\ 
    &= 0,
\end{aligned}
\end{equation}

where we have used anticommutation relations from eqn. (\ref{eqn: ACRs}) and the fact that $v_1a_i^\dagger=a_iv_i=0$. The $U(1)$ generator $\Lambda_{15}$ commutes with $\{\Lambda_1,...,\Lambda_8\}$ and is proportional to the electric charge generator defined as;
\begin{equation} \label{eq:Q1}
    Q_1 := \frac{1}{3}(a_1^\dagger{a_1}+a_2^\dagger{a_2}+a_3^\dagger{a_3}-3a_4^\dagger{a_4}) \hspace{1em} \in \mathfrak{su}(4),
\end{equation}
which assigns the correct electric charge to each state. One generation of fermions with $SU(3)_C\times U(1)_{em}$ gauge symmetry may therefore be represented in terms of the minimal left ideal $T_1$. The ideal basis elements can be uniquely identified with corresponding particles depending on how they transform under $SU(3)_C$ and $U(1)_{em}$. The $\mathbb{C}\ell(6)v_1$ terms represent the isospin-up states whereas the $\mathbb{C}\ell(6)a_4^\dagger{v_1}$ terms represent the isospin-down states. For example;
\begin{equation}
\begin{aligned}\relax
   [Q_1, a_1^\dagger v_1] &= \frac{1}{3}(a_1^\dagger{a_1}+a_2^\dagger{a_2}+a_3^\dagger{a_3}-3a_4^\dagger{a_4})(a_1^\dagger v_1) \\
     &-\frac{1}{3}(a_1^\dagger v_1)(a_1^\dagger{a_1}+a_2^\dagger{a_2}+a_3^\dagger{a_3}-3a_4^\dagger{a_4}) \\ 
 &= \frac{1}{3}(a_1^\dagger a_1)(a_1^\dagger v_1) \\
 &= +\frac{1}{3}a_1^\dagger v_1, \\
 [Q_1, a_1^\dagger a_4^\dagger v_1] &= \frac{1}{3}(a_1^\dagger{a_1}+a_2^\dagger{a_2}+a_3^\dagger{a_3}-3a_4^\dagger{a_4})(a_1^\dagger a_4^\dagger v_1) \\
 &-\frac{1}{3}(a_1^\dagger a_4^\dagger v_1)(a_1^\dagger{a_1}+a_2^\dagger{a_2}+a_3^\dagger{a_3}-3a_4^\dagger{a_4}) \\
 &= \frac{1}{3}(a_1^\dagger a_1 - 3a_4^\dagger a_4)(a_1^\dagger a_4^\dagger v_1) \\ 
 &= \frac{1}{3}(a_1^\dagger a_4^\dagger v_1 - 3a_1^\dagger a_4^\dagger v_1) \\ 
 &= -\frac{2}{3}a_1^\dagger a_4^\dagger v_1,
\end{aligned}
\end{equation}

where $a_1^\dagger v_1$ corresponds to a $+\frac{1}{3}$ charged anti-down quark while $a_1^\dagger a_4^\dagger v_1$ corresponds to a $-\frac {2}{3}$ charged anti-up quark. This construction is different to the construction in \cite{Furey2016}, which uses two $\mathbb{C}\ell(6)$ ideals, generated from $\mathbb{C}\otimes\mathbb{O}$, with $U(1)_{em}$ arising from the $\mathbb{C}\ell(6)$ number operator ($N:=\frac{1}{3}(a_1^\dagger{a_1}+a_2^\dagger{a_2}+a_3^\dagger{a_3})$). 

The $SU(3)$ generators above correspond to a subgroup of $Spin(6)$ generated from $\mathbb{C}\ell(6)$ bi-vectors. This same $SU(3)$ corresponds to a subgroup of the octonion automorphism group $G_2$, and in fact $SU(3)=G_2 \cap Spin(6)$ \cite{Dixon1994}. On the other hand, $Q_1$ defined above is not a subgroup of this $Spin(6)$ or $G_2$, and can only be defined within $Spin(8)$.

\section{$S_3$ as a generation symmetry}

We recall that $Aut(\mathbb{S})=G_2\times S_3$, and that $\mathbb{C}\ell(8)$ corresponds to the left multiplication algebra of $\mathbb{C}\otimes\mathbb{S}$. At the core of our approach is the proposal that the discrete automorphism group $S_3$ of $\mathbb{C} \otimes \mathbb{S}$ provides the algebraic source for three generations. Such a discrete generation symmetry is lacking in models based purely on division algebras. We now extend the representation of a single generation of fermions, with $SU(3) \times U(1)$ gauge symmetry, in terms of a single minimal left ideal of $\mathbb{C}\ell(8)$, to three generations. We do this by embedding the $S_3$ automorphisms of $\mathbb{C}\otimes\mathbb{S}$ into $\mathbb{C}\ell(8)$, and subsequently interpret this discrete group as a generation symmetry.

\subsection{Extending the $S_3$ discrete symmetry to $\mathbb{C}\ell(8)$}

The automorphism $\phi$ defined in eqn. (\ref{eqn: phi}) has a natural extension to $\mathbb{C}\ell(8)$; by letting the $\phi$ automorphisms act on the left actions $e_i$ instead of sedenion elements $s_i$. The $\epsilon$ automorphism likewise has an obvious embedding into $\mathbb{C}\ell(8)$: $e_i\xrightarrow{\epsilon}{e_i}$ for $i \in \{0,1,...,7\}$ and $e_8\xrightarrow{\epsilon}{-e_8}$ within $\mathbb{C}\ell(8)$\footnote{It then follows (see eqn. (\ref{eqn:e9})) that $e_j\xrightarrow{\epsilon}{-e_j}$, for $j \in \{9,...,15\}$.}.

In our previous paper \cite{Gresnigt2023}, we defined the extension of $\psi$ into $\mathbb{C}\ell(8)$ via the map;
\begin{eqnarray}
    e_i \xrightarrow{\psi_{old}}
    \begin{cases}
    -\frac{1}{2}e_i-\frac{\sqrt{3}}{2}e_{i+8} & i=\{1,...,7\},\\
    -\frac{1}{2}e_{i}+\frac{\sqrt{3}}{2}e_{i-8} & i=\{9,...,15\},\\
    e_i & i=\{0,8\}.
    \end{cases}
\end{eqnarray}

This definition does not allow for the construction of a $U(1)$ generator that assigns the correct electric charge to three generations of states. Furthermore, $e_i$, $\psi_{old}(e_i)$ and $\psi_{old}^2(e_i)$ are linearly dependent, satisfying;
\begin{eqnarray}
    e_i+\psi_{old}(e_i)+\psi_{old}^2(e_i)=0,
\end{eqnarray}
leading to an inevitable linear dependence between some states. 

Instead, here we propose a generalised extension of $\psi\in Aut(\mathbb{S})$ into $\mathbb{C}\ell(8)$ by defining\footnote{One way to motivate this new map is by defining the involution of a $\mathbb{C}\ell(8)$ element as $(e_i)^*\rightarrow-e_{i+8}e_8$. Inserting this involution into eqn. (\ref{eqn:Psi}) and converting the $s_i$ to $e_i$ then results in the new definition of $\psi$.};
\begin{equation}
e_i \xrightarrow{\psi}
\begin{cases}
    \frac{1}{4}e_i-\frac{\sqrt{3}}{4}e_ie_8+\frac{\sqrt{3}}{4}e_{i+8}-\frac{3}{4}e_{i+8}e_8  & i=\{0,...,7\},\\
    \frac{1}{4}e_i-\frac{\sqrt{3}}{4}e_ie_8-\frac{\sqrt{3}}{4}e_{i-8}+\frac{3}{4}e_{i-8}e_8 & i=\{8,...,15\}.
\end{cases}
\end{equation}
For example;
\begin{align}
    e_1 &\xrightarrow{\psi} \frac{1}{4}e_1-\frac{\sqrt{3}}{4}e_1e_8+\frac{\sqrt{3}}{4}e_{9}-\frac{3}{4}e_{9}e_8.
\end{align}

While the $e_{i+8}$ terms can be rewritten as $\mathbb{C}\ell(8)$ elements as per eqn. (\ref{eqn:e9}), it is more convenient to leave them as $e_{i+8}$. It can then be checked that $\psi^3(e_i)=e_i$, and that both $e_0$ and $e_8$ are invariant under $\psi$. The $\mathbb{C}\ell(8)$ maps $\epsilon$ and $\psi$ can then be seen to generate $S_3$.

We pause to mention that although many of the calculations that follow are straightforward, they are very tedious to carry out by hand. This is because, for example, the transformed basis vectors $\psi(e_i)$ involve ten terms once $e_{i+8}$ is rewritten as an element of $\mathbb{C}\ell(8)$ using eqn. (\ref{eqn:e9}). The authors have used Mathematica to verify the calculations that follow.

One can subsequently check that;
\begin{align} \label{eqn: Auto1}
    \psi(e_i)\psi(e_i)=\psi(e_i^2)&=-1, \\ \label{eqn: Auto2}
    \psi(e_i)\psi(e_j)+\psi(e_j)\psi(e_i)&=0,
\end{align}
for $i,j \in \{0,1,...,8\},\; i \neq j$, and likewise for $\epsilon$. These maps therefore extend to $\mathbb{C}\ell(8)$ homomorphisms \cite{harvey1990spinors}. Unlike in our previous paper, $e_i$, $\psi(e_i)$ and $\psi^2(e_i)$ are now linearly independent in $\mathbb{C}\ell(8)$.

\subsection{Including two additional generations using the order-three symmetry $\psi$}

Applying $\psi$ (and $\psi^2$) to $a_i$ and $a_i^\dagger$ generates two additional Witt bases;
\begin{equation}
b_i=\psi(a_i), \hspace{1em} b_i^\dagger=\psi(a_i^\dagger), \hspace{1em} c_i=\psi^2(a_i), \hspace{1em} c_i^\dagger=\psi^2(a_i^\dagger),\quad i=\{1,2,3,4\},
\end{equation}
satisfying the same fermionic anticommutation relations (eqn. (\ref{eqn: ACRs})) as our original Witt basis, with $\{a_i, a_i^\dagger\}$ replaced with $\{b_i, b_i^\dagger\}$ or $\{c_i, c_i^\dagger\}$. We can therefore construct two additional minimal left ideals;
\begin{equation} 
T_2 = (r_0^{'}+r_j^{'}b_j^\dagger+r_{jk}^{'}b_j^\dagger{b_k^\dagger}+r_{jkl}^{'}b_j^\dagger{b_k^\dagger}b_l^\dagger+r_{1234}^{'}b_1^\dagger{b_2^\dagger}b_3^\dagger{b_4^\dagger})v_2,
\end{equation}
\begin{equation} 
T_3 = (r_0^{''}+r_j^{''}C_j^\dagger+r_{jk}^{''}c_j^\dagger{c_k^\dagger}+r_{jkl}^{''}c_j^\dagger{c_k^\dagger}c_l^\dagger+r_{1234}^{''}c_1^\dagger{c_2^\dagger}c_3^\dagger{c_4^\dagger})v_3,
\end{equation}
where $v_2=\psi(v_1)$, $v_3=\psi^2(v_1)$, $j,k,l \in \{1,2,3,4\}$, $j\neq{k}\neq{l}$ and $r_0^{'},r_j^{'},r_{jk}^{'},r_{jkl}^{'},r_{1234}^{'}$, $r_0^{''},r_j^{''},r_{jk}^{''},r_{jkl}^{''},r_{1234}^{''}$ are complex coefficients. Applying $\psi$ to $T_3$ returns $T_1$, and so $\psi$ permutes between three minimal ideals.

\subsection{Generation invariant gauge symmetries}

We wish to identify the minimal left ideals $T_2$ and $T_3$ with the second and third generation of fermions. As with the first ideal $T_1$, not all of $Spin(8)$ preserves the ideals via commutation, only a $U(4)$ subgroup does. However, the $SU(4)$ generators defined in eqn. (\ref{eq:SU(4)}) do not transform the basis states of $T_2$ and $T_3$ as $1\oplus4\oplus6\oplus\overline{4}\oplus1$. 

In order to identify the gauge symmetries in our model, we impose two conditions:
\begin{enumerate}
    \item The gauge symmetries must preserve the minimal left ideals under commutation.
    \item The gauge symmetries must be invariant under (commute with) the action of $S_3$.
\end{enumerate}
It can be checked that $\Lambda_i=\psi(\Lambda_i),\; i=\{1,...,8\}$, and that the ideals $T_2$ and $T_3$ likewise transform as one fermion generation under the action of this $SU(3)$. For example;
\begin{equation}
\begin{aligned}\relax
 [\Lambda_1, b_1^\dagger b_3^\dagger b_4^\dagger v_2] &= (-a_2^\dagger a_1 - a_1^\dagger a_2)(b_1^\dagger b_3^\dagger b_4^\dagger v_2)-(b_1^\dagger b_3^\dagger b_4^\dagger v_2)(-a_2^\dagger a_1 - a_1^\dagger a_2)\\
 &=(-b_2^\dagger b_1 - b_1^\dagger b_2)(b_1^\dagger b_3^\dagger b_4^\dagger v_2)-(b_1^\dagger b_3^\dagger b_4^\dagger v_2)(-b_2^\dagger b_1 - b_1^\dagger b_2) \\
 &= -b_2^\dagger b_3^\dagger b_4^\dagger v_2.
\end{aligned}
\end{equation}
The gauge symmetry $SU(3)_C$ identified in the single-generation construction therefore extends to three generations.

On the other hand, the $U(1)$ generator $Q_1$ turns out not to be invariant under $S_3$. Nonetheless, it is possible to construct an $S_3$-invariant $U(1)$ generator as the sum of three individual $U(1)$ generators;

\begin{equation} 
Q:=\frac{1}{3}(Q_1+\psi(Q_1)+\psi^2(Q_1)).    
\end{equation}

This new $U(1)$ commutes with $SU(3)$. Although $Q$ preserves the ideal $T_1$ (and $T_2$, $T_3$), it no longer assigns the electric charge corresponding to both isospin-up and isospin-down states. Instead, $Q$ assigns eigenvalues to the basis states of $T_1$ corresponding to two copies of isospin-down states (and likewise for $T_2$ and $T_3$). For example, both $a_1^\dagger a_2^\dagger v_1$ and $a_1^\dagger a_2^\dagger a_4^\dagger v_1$ correspond to the same colour down quark (their transformation under $SU(3)$ is identical);

\begin{equation}
   [Q, a_1^\dagger a_2^\dagger v_1] = -\frac{1}{3} a_1^\dagger a_2^\dagger v_1, \qquad
    [Q, a_1^\dagger a_2^\dagger a_4^\dagger v_1] = -\frac{1}{3} a_1^\dagger a_2^\dagger a_4^\dagger v_1.
\end{equation}

We will address this issue shortly. The remaining $SU(4)$ generators could likewise be generalised to be $S_3$-invariant, by defining;

\begin{equation} 
\Delta_i:=\frac{1}{3}(\Lambda_i+\psi(\Lambda_i)+\psi^2(\Lambda_i)), \quad i=9,...,14.
\end{equation}

However, one finds that the minimal ideals are not preserved under commutation with these generators. We therefore exclude them from being part of any viable gauge symmetries.

\subsection{Including isospin-up states using the order-two symmetry $\epsilon$}
To include the isospin-up states for the first generation, we utilise the order-two symmetry $\epsilon$. Its action on the three Witt bases is as follows;
\begin{alignat}{4} \notag
    \epsilon(a_i)&=a_i, \hspace{1em} \epsilon(a_4)&&=-a_4^\dagger, \hspace{1em} \epsilon(a_i^\dagger)&&&=a_i^\dagger, \hspace{1em} \epsilon(a_4^\dagger)=-a_4, \\ 
    \epsilon(b_i)&=c_i, \hspace{1em} \epsilon(b_4)&&=-c_4^\dagger, \hspace{1em} \epsilon(b_i^\dagger)&&&=c_i^\dagger, \hspace{1em} \epsilon(b_4^\dagger)=-c_4, \\ \notag
    \epsilon(c_i)&=b_i, \hspace{1em} \epsilon(c_4)&&=-b_4^\dagger, \hspace{1em} \epsilon(c_i^\dagger)&&&=b_i^\dagger, \hspace{1em} \epsilon(c_4^\dagger)=-b_4, 
\end{alignat}
with $i=1,2,3$ \footnote{As an example; \begin{align} \notag \epsilon(b_1^\dagger)&=\epsilon(\frac{1}{4}e_1-\frac{\sqrt{3}}{4}e_1e_8+\frac{\sqrt{3}}{4}e_9-\frac{3}{4}e_9e_8+\frac{i}{4}e_5-\frac{\sqrt{3}i}{4}e_5e_8+\frac{\sqrt{3}i}{4}e_{13}-\frac{3i}{4}e_{13}e_8) \\ \notag &=(\frac{1}{4}e_1+\frac{\sqrt{3}}{4}e_1e_8-\frac{\sqrt{3}}{4}e_9-\frac{3}{4}e_9e_8+\frac{i}{4}e_5+\frac{\sqrt{3}i}{4}e_5e_8-\frac{\sqrt{3}i}{4}e_{13}-\frac{3i}{4}e_{13}e_8)=c_1^\dagger. \end{align}}. Applying $\epsilon$ to $T_1$ produces a second minimal left ideal, defined as $S_1 := \epsilon(T_1)$, built on the primitive idempotent $v_1':=\epsilon(v_1)$. The symmetry generator $Q$ then identifies the basis states of this complementary ideal as (two copies of) isospin-up states. For example, both $a_1^\dagger a_2^\dagger v_1^{'}$ and $a_1^\dagger a_2^\dagger a_4 v_1^{'}$ correspond to the same colour anti-down quark;
\begin{equation} 
    [Q, a_3^\dagger v_1^{'}] = \frac{1}{3} a_3^\dagger v_1^{'}, \qquad
    [Q, a_3^\dagger a_4 v_1^{'}] = \frac{1}{3} a_3^\dagger a_4 v_1^{'}.
\end{equation}

The fact that each isospin-down state is represented twice in $T_1$ indicates there is an additional degree of freedom that can be included, most likely chirality. Since this is not the focus of the present paper, we will reduce $T_1$ (and $S_1$) to its even semi-spinor via the projector $\rho_{+}=\frac{1}{2}(1+ e)$, where $e:=e_1e_2e_3e_4e_5e_6e_7e_8$ is the $\mathbb{C}\ell(8)$ pseudoscalar. Explicitly, the even semi-spinors $T_1^+:=\rho_+T_1$ and $S_1^+:=\rho_+S_1$ are given by;

\begin{equation}
\begin{split}
    T_1^{+} = &\Bigg(
    \begin{array}{c}
    \overline{\nu_e} \\
    + d^r a_1^\dagger a_2^\dagger + d^g a_1^\dagger a_3^\dagger + d^b a_2^\dagger a_3^\dagger \\
    + \overline{u^b} a_1^\dagger a_4^\dagger + \overline{u^g} a_2^\dagger a_4^\dagger + \overline{u^r} a_3^\dagger a_4^\dagger \\
    + e^- a_1^\dagger a_2^\dagger a_3^\dagger a_4^\dagger
    \end{array}\Bigg) v_1,
\end{split}
\end{equation}

\begin{equation} 
\begin{split}
   S_1^+ = &\Bigg(
    \begin{array}{c}
     \nu_e \\
    + u^r a_1^\dagger a_2^\dagger + u^g a_1^\dagger a_3^\dagger + u^b a_2^\dagger a_3^\dagger \\
    + \overline{d^b} a_1^\dagger a_4 + \overline{d^g} a_2^\dagger a_4 + \overline{d^r} a_3^\dagger a_4 \\
    + e^+ a_1^\dagger a_2^\dagger a_3^\dagger a_4 
    \end{array}\Bigg)  v_1',
\end{split}
\end{equation}
where the semi-spinors' suggestively named coefficients indicate how they transform. The $S_3$-invariant $U(1)$ generator $Q$ now assigns the correct electric charge to all the states, with $T_1^+$ containing the isospin-down states, and $S_1^{+}$ the isospin-up states. Explicitly, for $T_i^{+}$ and $S_i^+$, respectively, the eigenvalues are;

\begin{align} \notag
    &0 \\ \notag
    -\frac{1}{3} \hspace{1em} -&\frac{1}{3} \hspace{1em} -\frac{1}{3} \\ \notag
    -\frac{2}{3} \hspace{1em} -&\frac{2}{3} \hspace{1em} -\frac{2}{3} \\ \notag
    -&1
 \end{align}

 \begin{align} \notag
    &0 \\ \notag
    \hphantom{-}\frac{2}{3} \hspace{1em} \hphantom{-}&\frac{2}{3} \hspace{1em} \hphantom{-}\frac{2}{3} \\ \notag
    \hphantom{-}\frac{1}{3} \hspace{1em} \hphantom{-}&\frac{1}{3} \hspace{1em} \hphantom{-}\frac{1}{3} \\ \notag
    &1
 \end{align}

Acting with the order-three symmetry $\psi$ then permutes between the even semi-spinors of the remaining two generations\footnote{Note that the order-two symmetry $\epsilon$ acting on a second generation (semi-)spinor results in a third generation (semi-)spinor and vice versa. For example, $\epsilon(T_2)=S_3$, or more explicitly $\epsilon(b^{\dagger}_1b^{\dagger}_4v_2)=-c_1^{\dagger}c_4v_3'$.}. One generation of fermions with unbroken $SU(3)_C\times U(1)_{em}$ gauge symmetry can therefore be represented in terms of two $\mathbb{C}\ell(8)$ semi-spinors, related via the $S_3$ order-two symmetry $\epsilon$. Two additional generations of states transforming identically to the first under the same symmetry are obtained by applying the $S_3$ order-three symmetry $\psi$ to the first generation of states. In this way, both $S_3$ generators are given a physical interpretation. Whereas $\epsilon$ interchanges isospin-down and isospin-up states, $\psi$ permutes between generations. In contrast to \cite{Gresnigt2023}, all three generations of states are linearly independent in this construction, a desirable feature.

\section{Discussion}

In this paper we have demonstrated that it is possible to algebraically represent three linearly independent generations of fermions with $SU(3)_C\times U(1)_{em}$ gauge symmetry within $\mathbb{C}\ell(8)$. Central to our construction is the idea that an $S_3$ discrete symmetry, arising from the automorphism group of $\mathbb{S}$, $Aut(\mathbb{S})=G_2\times S_3$, is the algebraic source for the existence of three generations. $\mathbb{C}\ell(8)$ corresponds to the multiplication algebra of $\mathbb{C}\otimes\mathbb{S}$. Our generation symmetry then corresponds to an embedding of the $S_3$ automorphisms of $\mathbb{S}$ into $\mathbb{C}\ell(8)$. In the resulting model, $S_3$ symmetry permutes between the three generations, and interchanges between isospin-down and isospin-up states. 

Our work here builds on previous attempts \cite{Gresnigt2023, Gillard2019} to use the sedenions to construct an algebraic model of three generations. Although these earlier works were able to describe three generations with $SU(3)_C$ gauge symmetry, they were unable to include the remaining unbroken symmetry $U(1)_{em}$. Additionally, the three generations of states were not entirely linearly independent. Our present model is able to incorporate both unbroken gauge symmetries by generalising the embedding of the $S_3$ automorphisms of $\mathbb{S}$ into $\mathbb{C}\ell(8)$. As an unexpected byproduct of this construction, the three generations of states are now linearly independent. 

One generation is represented in terms of two semi-spinors, obtained from two minimal left ideals of $\mathbb{C}\ell(8)$, related via the order-two symmetry of $S_3$. One semi-spinor contains the isospin-down states, whereas the other the isospin-up states. Applying the order-three $S_3$ symmetry then produces two additional pairs of semi-spinors, linearly independent to the first, to represent the remaining two generations. 

The gauge symmetries are identified as the unitary symmetries that both preserve the semi-spinors under commutation and are invariant under $S_3$. Whereas the $SU(3)_C$ symmetry constructed for the first generation is inherently invariant under $S_3$, an $S_3$-invariant $U(1)_{em}$ consists of three separate $U(1)$ symmetries, one associated with each generation, which individually are not $S_3$-invariant. 

The next step to develop this model further is to see how electroweak symmetry $SU(2)_L\times U(1)_Y$ may be included. As with existing single-generation models, this will likely involve enlarging the algebra to $\mathbb{C}\ell(10)$ \cite{Furey2016, Gresnigt2020, Todorov2022}. One way to achieve this is to include a factor of the quaternions $\mathbb{H}$. The left multiplication algebra of $\mathbb{C}\otimes\mathbb{H}\otimes\mathbb{S}$ would then be $\mathbb{C}\ell(10)$\footnote{This could be further extended to $\mathbb{C}\ell(12)$ if both the left and right actions of $\mathbb{H}$ onto $\mathbb{C}\otimes\mathbb{H}\otimes\mathbb{S}$ are considered.}. 

Electroweak symmetry breaking occurs when the Higgs field acquires a non-zero vacuum expectation value, breaking the SM gauge group $SU(3)_C \times SU(2)_L \times U(1)_{Y}$ to the unbroken $SU(3)_C\times U(1)_{em}$. One consequence of this is the resulting massless bosons, the photon and gluons, at low energies. In contrast, the bosons involved in the broken symmetries, the $W^\pm$ and Z bosons, acquire mass at low energies. We do not, at this point, speculate how this mass generation arises algebraically in our model (or an extension thereof).

It is worth noting that various $S_3$ extensions to the SM have been considered in the literature, particularly in relation to Higgs, neutrinos, and flavour physics \cite{Kubo2004, Kubo2005, Kubo2004a,Hernandez2021, Dong2012, Vien2014}. This makes us hopeful that our model, once the weak interaction is included, will be able to describe additional features such as neutrino oscillations and quark mixing, as well as make phenomenological predictions, something that is lacking in current division algebra based models \footnote{Some recent papers attempt to address these issues. Patel and Singh attempt to derive the CKM matrix parameters from the exceptional Jordan algebra, whereas Tang and Tang look at sedenions \cite{Patel2023,Tang2023}.}.

Other discrete groups have also been proposed as extensions to the SM. The alternating group $A_4$ has been proposed as a symmetry extension in \cite{Ding2021, Novichkov2019, Chen2021, Altarelli2010}, $S_4$ symmetry extensions have been proposed in \cite{King2021, Wang2019, Bjoerkeroth2017, Thapa2023}, while there have also been dihedral group $D_4$ symmetry extensions proposed in \cite{Ishimori2013, Vien2018}. These proposals aim to address the problem of the mass and mixing hierarchies observed in the quark and lepton sectors, particularly the neutrino mixing pattern and the small masses of neutrinos. We note that these discrete symmetry groups do not arise as automorphism groups of Cayley-Dickson algebras. Our model presented here, based on sedenions, is therefore incompatible with such proposed extensions to the SM, as it is uniquely $S_3$ that arises  as a discrete automorphism group of the algebra. 

Finally, several authors have proposed that triality might account for the existence of three generations \cite{perelman2021jordan,dubois2016exceptional1,dubois2019exceptional2,boyle2020standard2}. Triality corresponds to an outer automorphism of $Spin(8)$ that permutes the vector and two spinor representations, all of dimension eight. It is known that three conjugacy classes of $Spin(7)$ subgroups in $Spin(8)$ are permuted by triality \cite{Varadarajan2001}. A unique selection of one $Spin(7)$ subgroup from each conjugacy class, with a common intersection of $G_2$, hints at the possibility of representing three generations of colour states, where each generation aligns with a distinct $Spin(7)$ subgroup. This triality automorphism stabilises the $G_2$ subgroup within $Spin(7)$ while rotating the remaining group elements isoclinically \cite{Lounesto1986}. It would be worthwhile investigating if the $S_3$ generation symmetry considered here can be identified with the triality automorphism. This is currently under investigation.

\subsection*{Acknowledgments}
NG would like to thank Alessio Marrani, Carlos Perelman, and Tejinder Singh for many insightful discussions related to the ideas presented in this paper.

\appendix

\section{Sedenion multiplication table}

The sedenion multiplication table can be seen in Table \ref{sedenions_table}.

\begin{table}[h!]
\caption{Sedenion multiplication table where $s_i$ is represented by $i$.}
\label{sedenions_table}
\[
\setlength{\arraycolsep}{2pt}
\renewcommand{\arraystretch}{1}
\begin{array}{c|*{16}{C}}
    &  0  &  1  &  2  &  3  &  4  &  5  &  6  &  7  &  8  &  9  & 10 & 11 & 12 & 13 & 14 & 15 \\
    \hline
 0  &  0  &  1  &  2  &  3  &  4  &  5  &  6  &  7  &  8  &  9  & 10 & 11 & 12 & 13 & 14 & 15 \\
 1  &  1  & -0  &  3  & -2  &  5  & -4  & -7  &  6  &  9  & -8  &-11 & 10 &-13 & 12 & 15 &-14 \\
 2  &  2  & -3  & -0  &  1  &  6  &  7  & -4  & -5  & 10 & 11 & -8  & -9  &-14 &-15 & 12 & 13 \\
 3  &  3  &  2  & -1  & -0  &  7  & -6  &  5  & -4  & 11 &-10 &  9  & -8  &-15 & 14 &-13 & 12 \\
 4  &  4  & -5  & -6  & -7  & -0  &  1  &  2  &  3  & 12 & 13 & 14 & 15  & -8  & -9  & -10 &-11 \\
 5  &  5  &  4  & -7  &  6  & -1  & -0  & -3  &  2  & 13 & -12 & 15 &-14 & 9 &  -8  & 11 & -10 \\
 6  &  6  &  7  &  4  & -5  & -2  &  3  & -0  & -1  & 14 & -15 & -12 & 13 & 10 &-11 &  -8  & 9 \\
 7  &  7  & -6  &  5  &  4  & -3  & -2  &  1  & -0  & 15 & 14 & -13 & -12 & 11 & 10 & -9 &  -8 \\
 8  &  8  & -9  &-10 &-11 &-12 &-13 &-14 &-15 & -0  &  1  &  2  &  3  &  4  &  5  &  6  &  7 \\
 9  &  9  &  8  &-11 & 10 & -13 & 12 & 15 & -14 & -1 & -0  & -3  &  2  & -5  &  4  &  7  & -6 \\
10 & 10 & 11 &  8  & -9  &-14 & -15 & 12 & 13 & -2  &  3  & -0  &  -1  & -6  & -7  &  4  &  5 \\
11 & 11 &-10 &  9  &  8  &-15 & 14 & -13 & 12 & -3  & -2  &  1  & -0  & -7  &  6  & -5  &  4 \\
12 & 12 & 13 & 14 & 15 &  8  &  -9  &-10 & -11 &  -4  &  5  &  6  &  7  & -0  &  -1  &  -2  &  -3 \\
13 & 13 &-12 & 15 &-14 &  9  & 8  & 11 & -10 & -5  &  -4  &  7  & -6  &  1  & -0  &  3  & -2 \\
14 & 14 &-15 &-12 & 13 & 10 & -11 &  8  &  9  & -6  &  -7  & -4  &  5  &  2  & -3  & -0  &  1 \\
15 & 15 & 14 &-13 &-12 & 11 & 10 &  -9  &  8  & -7  & 6  & -5  & -4  &  3  &  2  & -1  & -0 \\
\end{array}
\]
\end{table}

\printbibliography

\end{document}